

\magnification=1200
\baselineskip=16pt
\pageno=1
\hskip 10cm
ICN-UNAM 95010
\vskip2pc
\centerline{\bf  THE CONSTRAINTS IN SPHERICALLY SYMMETRIC}
\centerline{\bf  GENERAL RELATIVITY III}
\vskip2pc
\centerline{\bf IDENTIFYING THE CONFIGURATION SPACE:}
\centerline{\bf  $J\ne 0$}
\vskip2pc
\centerline {\bf Jemal Guven$^{(1)}$ and Niall \'O Murchadha$^{(2)}$ }
\vskip1pc
\it
\centerline {$^{(1)}$Instituto de Ciencias Nucleares}
\centerline {Universidad Nacional Aut\'onoma de M\'exico}
\centerline {A. Postal 70-543. 04510 M\'exico, D. F., MEXICO} \rm
\centerline{(guven@roxanne.nuclecu.unam.mx)}
\vskip2pc
\it
\centerline {$^{(2)}$ Physics Department}
\centerline {University College Cork}
\centerline {Cork, IRELAND} \rm
\centerline{(niall@iruccvax.ucc.ie)}
\vfill\eject
\centerline{\bf Abstract}
\vskip1pc
{\leftskip=1.5cm\rightskip=1.5cm\smallskip\noindent
We continue our examination of the constraints in
spherically symmetric general relativity. We extend to
general configurations with $J\ne 0$ the analysis
of II which treated a moment of time symmetry.
We exploit the one parameter family of foliations introduced in I
which are linear and homogeneous in the extrinsic curvature to
characterize apparent horizons and spatial singularities in the
initial data. In particular, we demonstrate that these
characterizations do not depend sensitively on the
foliation.\smallskip}
\vskip2pc

\vfill
\eject

\noindent{\bf 1. INTRODUCTION}
\vskip1pc

This is the third paper in a series in which we examine the general
features of the constraints in general relativity under the assumption that
the spatial geometry is spherically symmetric and possesses just one
asymptotically flat region [1,2]. In paper II, we focused on solutions
of the constraints which occur when the extrinsic curvature $K_{ab}$
momentarily vanishes (MSCs). As such, we did not need to address the
issue of fixing the foliation. In this paper, we extend this analysis to
incorporate a non-vanishing extrinsic curvature. Unless one sets
out to be difficult, this corresponds to a non-vanishing flow of
matter, $J$.

The introduction of extrinsic curvature complicates the analysis
substantially. The advantage of having dealt separately with moment of
time symmetry configurations in paper II is that we can focus here on
the physical feedback on the spatial geometry introduced by extrinsic
curvature. The important
point is that the solutions of the constraints, as
well the relationships between the global measures of the energy and
the dimensions of the support of matter which we exploit to characterize
horizons and singularities, are not sensitively dependent on the gauge
fixing the foliation.

To fix this foliation we
implement explicitly one of the gauges parametrized by $\alpha$,
linear and homogeneous in the extrinsic curvature, introduced in paper I.
These gauges will serve to set in context our understanding of
the constraints when the initial data is momentarily static [2].
It was shown in paper I that the allowed values assumed by the
parameter correspond to all tangent vectors lying within the
superspace lightcone. One of these gauges is the maximal slicing gauge.
Another is the polar gauge. Among the attractive properties of all such gauges
is that they  foliate flat spacetime by flat spatial hypersurfaces. When
the momentum
constraint is satisfied, the extrinsic curvature is linear in $J$, albeit
in a non-local way. In this way the extrinsic curvature of the hypersurface
responds directly to the (radial) movement of matter on it.
Another possibility, which is motivated by the introduction of
the optical scalars as canonical variables on the phase space, is to treat
the trace of the extrinsic curvature, $K$, itself as
an independent datum along with energy density, $\rho$, and
$J$. While this is a legitimate gauge,
it is not a usual one unless $K=0$, for if $K\ne 0$, the extrinsic curvature
cannot adjust itself to the movement of matter. For this reason it
does not correspond to our physical expectations and, therefore,
we do not consider this possibility further here.

As we did in paper II, we will focus again on the identification of
the global structures that characterize non-trivial
geometries ---  apparent horizons and singularities.

The identification of apparent horizons is complicated by the
fact that these physical landmarks no longer coincide with the extremal
surfaces of the spatial geometry as they do at a
moment of time symmetry: the movement of matter generates extrinsic
curvature, thereby affecting how the spatial geometry is
embedded in spacetime which, in turn, determines the lightcone
structure on the surface.

We begin in sect.2 with a discussion of the generic analytic
structure of the constraints.
We derive a spatial diffeomorphism invariant analytic expression
for the behavior of the geometry in the neighborhood of a
generic singularity. Generally, the singularities of the
three-geometry consistent with the constraints will be more severe
than those which are possible at a moment of time symmetry.
If, however, the movement of matter is tuned so that the extrinsic
curvature vanishes as the singularity is approached,
the strength of the singularity will be determined entirely by the
quasi-local mass (QLM), exactly as it
was at a moment of time symmetry [3]. This tuning
corresponds to an integrability condition on the current.
If, in addition, the tuning
is refined such that the QLM also vanishes as we approach the singularity the
curvature singularity disappears and the spatial geometry
pinches off in a regular way. This latter integrability condition
involving the QLM is completely analogous to the integrability
condition we encountered at a moment of time symmetry.
Regularity at the singularity is, of course, precisely the condition
that the interior be a regular closed universe. If the matter fields carry
conserved charges these will, in their turn, have integrability conditions
associated with them. Viewed this way, regular closed universes
appear to be very special universes [4].

In paper I, we represented the configuration space of the
spherically symmetric theory by bounded
closed trajectories on the optical scalar plane.
In sect.3 we examine these trajectories in vacuum.
We discover that any trajectory that finds itself
outside a proper subset of this domain is necessarily singular.
While the detailed structure of this good subset depends on the specific gauge
choice we make to determine the slicing, we find many features that are
independent of the slicing. For example, on the boundary of the good subset
we find two unstable fixed points. These correspond to the situation where the
exterior spatial geometry neither collapses to a singularity nor expands to be
asymptotically flat. Instead it becomes a semi-infinite cylinder of fixed
radius.

We follow paper II by establishing global necessary and sufficient
conditions for the occurence of apparent horizons and
singularities. These conditions are framed in terms of inequalities which
relate some appropriate measure of the material energy content
on a given support to a measure of its volume. The challenge is
to identify useful measures in both cases.
In analogy to the total material energy
$M$ (defined in paper I), we can introduce the total momentum, $P$,
corresponding to the integrated material current over the proper
spatial volume. The sufficiency criteria for the formation of a future
(past) apparent horizon can be cast in a form which is a
straightforward generalization of the moment of time symmetry
inequality: if the difference (sum) of the material energy and
the material momentum exceeds some universal constant times
the proper radius, $\ell_0$, of the distribution, the geometry
will possess a future (past) apparent horizon.
The corresponding constant for singularities  is larger but the
inequality does not involve $P$. As we found at a MSC a more appropriate
measure of the material energy for casting the necessary criteria is the
maximum value of the energy density of matter, $\rho_{\rm Max}$. The
obvious generalization is the sum $\rho_{\rm Max}
+ J_{\rm Max}$. However, we find that that the inequality is not
symmetrical under interchange of $\rho$ and $J$. If the dominant
energy condition is satisfied, however, we can however cast the
inequalities in the momentarily static form:
if $(\rho_{\rm Max}+J_{\rm Max})\ell_0^2 <$
some constant, the distribution of matter does will not possess a
singularity for one constant and an apparent horizon for some
other smaller constant. These inequalities are new.

In the treatment by Bizon, Malec and \'O Murchadha (BM\'OM),
and more recently by Malec and \'O Murchadha of the
sufficiency conditions, the slicing of spacetime was
always assumed to be maximal with $K=0$ [5,6]. If these inequalities
are to be interpreted physically, they should, at least qualitatively,
be reproduced in other gauges. We note that we never needed to address this
question in paper II because the notion of a moment of time symmetry is
gauge independent. We examine the sensitivity of these
inequalities on the value of $\alpha$ appearing in the
gauge condition. Not surprisingly, we find that
the strength of the corresponding inequality does depend on the
foliation gauge but {\it not} in any significant way
so long as we are not close to the lightcone in superspace.

Unfortunately, unlike in our examination of the constraints at a moment
of time symmetry where we
could fall back on the piecewise-constant density models, we enjoy no such
exactly solvable standbys here.  Even the analogue of the constant density
star proves to be analytically  intractable when $J\ne 0$. It is not
surprising therefore that it is far more difficult to identify sharp
inequalities than it was at a moment of time symmetry. Much of our effort
is spent bootstrapping on moment of time symmetry inequalities.

There is one extremely useful exactly solvable model consisting of
a moving shell. We exploit this to speculate about the likely form of
a possible generalization to $J\ne 0$ of the lower bound
on the binding energy derived by BM\'OM.

\vskip2pc
\noindent{\bf 2. THE CONSTRAINTS}
\vskip1pc

In this section we examine various general features of the
constraints when $K_{ab}\ne 0$ in a manner which parallels,
wherever applicable, the treatment in paper II.
We recall that the constraints are given by

$$K_R\left[K_R+2K_{\cal L}\right]-
{1\over R^2}\Big[2 \left(R R^\prime \right)^\prime -R^{\prime 2}
-1 \Big]=8\pi \rho\eqno(2.1a)$$
and

$$K_R^\prime + {R^\prime \over R}(K_R-K_{\cal L})=4\pi J\,,\eqno(2.1b)$$
where the line element on the spatial geometry is parametrized

$$ds^2= d\ell^2+R^2 d\Omega^2\,,\eqno(2.1c)$$
and we have expanded the extrinsic curvature ($n^a$ is the
outward pointing unit normal to the two-sphere of fixed $\ell$),

$$K_{ab}= n_a n_b K_{\cal L} + (g_{ab}-n_a n_b)K_R\,.\eqno(2.1d)$$
All derivatives are with respect to the proper radius of the
spherical geometry, $\ell$.  The spatial geometries we
consider consist of a single asymptotically flat
region with a regular center, $\ell=0$. The appropriate boundary
condition on the metric at $\ell=0$ is then

$$R(0)=0\,.\eqno(2.2)$$
We recall that $R^\prime(0)=1$ if the geometry is regular at this point.
We assume that both $\rho$ and $J$ are appropriately bounded
functions of $\ell$ on some compact support. A
non-singular asymptotically flat solution defined for all
$\ell\ge 0$ will not, however, always exist for every specification of $\rho$
and $J$. Our task is to understand what can go wrong.

To solve the constraints classically, we need to implement
a foliation gauge. This involves some spatial scalar function of
the extrinsic curvature tensor. In a spherically symmetric geometry,
this tensor has only two independent components. The foliation either
fixes one of these or relates it functionally to the second.
Modulo the gauge, the momentum constraint can be
solved for this other component. The extrinsic curvature
is then completely determined by the sources.

In paper I, we introduced the one-parameter family
of gauges, defined for each $\alpha$ by

$$K_{\cal L}+\alpha K_R=0\,.\eqno(2.3)$$
We showed that the momentum constraint can be solved uniquely in
terms of the radial flow of matter, $J$, as follows (Eq.(3.3) of I)

$$ K_R = {4\pi\over R^{1+\alpha}}\int_0^{\ell} d\ell R^{1+\alpha} J
\,.\eqno(2.4)$$
Whenever $0.5<\alpha<\infty$ the gauge is valid everywhere and
displays the correct asymptotically flat falloff outside the support
of $J$ if the geometry is non-singular.
When Eqs.(2.3) and (2.4) are substituted into Eq.(2.1),
we obtain a second order singular non-linear integro-ODE for $R$.\footnote *
{It is possible to rewrite Eqs.(2.1) so that they can be differentiated
once to yield a {\it local} third order singular ODE modulo Eqs.(2.3) and
(2.4). However, it is not particularly illuminating to cast them this way.}
Subject to the boundary condition, (2.2), the solution is
uniquely determined. Not only is the extrinsic curvature
completely determined by the material sources, so also is the
spatial geometry. There are no independent gravitational
degrees of freedom.

We note that in the gauge Eq.(2.3), the spatial geometry
does not depend on the global sign of $J$.

We saw in paper I that if $K_R$ is regular at the origin
then it must also vanish there. In fact, in the neighborhood of
$\ell=0$, $R\sim \ell$, so that

$$K_R\sim 4\pi {J(0)\over 2+\alpha} \ell\,.\eqno(2.5)$$
To determine the $n^{\rm th}$ derivative of $R$ at $\ell=0$
we need to differentiate Eq.(2.1) $n-1$ times. A consequence of
the vanishing of $K_R(0)$  is that $J$ will only show up at order five ---
two orders behind $\rho$ (see paper II). The behavior of the
metric at the origin is clearly not sensitive to the current flowing there.

It is instructive to also examine the values
assumed by the optical scalars in the neighborhood of $\ell=0$.
Recall that [1,7]

$$\eqalign{\omega_+ =& 2\Big[R^\prime+ R K_R\Big]\,,\cr
\omega_- =& 2\Big[R^\prime- R K_R\Big]\,,\cr}\eqno(2.6a,b)$$
We can combine Eq.(2.4) of paper II and (2.5) to obtain

$$\omega_\pm \sim 2 - {8\pi\over 3}(\rho (0) \mp
{3\over 2+\alpha}J(0) ) \ell^2 \,.\eqno(2.7)$$
Suppose that the dominant energy condition (DEC) $\rho\ge |J|$ is satisfied.
If $\alpha\ge 1$, then $\omega_\pm \le 2$. If, however,
$\alpha<1$ this is not the case. This demonstrates explicitly that
the inequalities (6.2a) and (6.2b) in paper I
cannot generally be relaxed to
the $K=0$ value. We note also that Eq.(2.7) implies

$$\omega_+\omega_- \le (2 - {8\pi\over 3}[\rho (0)\ell^2])^2
- \left({8\over 2+\alpha}\right)^2
\pi^2 J(0)^2 \ell^4 \le 4 \,,$$
which is consistent with the inequality (5.2) in paper I
for all values of $\alpha$. Note also that the absolute
maximum of the product $\omega_+\omega_-$
obtains at the boundary values $\ell=0$ and $\ell=\infty$
and it is also the flat space value. When $K=0$, this is also true of both
$\omega_+$ and $\omega_-$. In general, the absolute maximum of neither
need occur at these points.

\vskip1pc
\noindent{ 2.1 The Quasi-Local Mass}
\vskip1pc

As we found in paper II in a simpler context, the definition of the
quasi-local mass can be exploited to provide an extremely useful
first integral of the constraints. We
recall that (Eq.(4.7) of paper I):

$$m = {R\over 2} \Big(1- R^{\prime2}\Big) + {1\over 2} K_R^2 R^3\,.
\eqno(2.8)$$
Modulo the constraints (Eq.(4.8$^\prime)$ of paper I),
and the boundary condition, (2.2),

$$m=4\pi \int_0^{\ell} d\ell R^2\left[\rho R^\prime  + J R K_R \right]
\,.\eqno(2.9)$$
If the geometry is non-singular, $m$ is positive everywhere,
coinciding in the limit $\ell\to \infty$ with the
ADM mass, $m_\infty$. Eqs.(2.8) and (2.9) are gauge invariant.
To exploit Eqs.(2.8) and (2.9) to solve the constraints,
we substitute the solution of the momentum constraint (2.4)
in the gauge (2.3) into Eqs.(2.8) and (2.9).

Note that the leading spatial derivative in Eq.(2.8) is $R^\prime$.
Outside the sources, $m$ is constant while $K_R \sim C/ R^{1+\alpha}$.
{}From a functional point of view, Eq.(2.8) is identical to the
energy integral in classical mechanics.
To exploit this analogy, we recast Eq.(2.8) for all $\ell$ as follows:

$$R^{\prime2}=1-{2m\over R} + K_R^2 R^2\,.\eqno(2.8^\prime)$$
Now, formally at least,

$$\ell =
\int_0^R  {dR\over \sqrt{1-{2m\over R} + K_R^2 R^2}}\,,$$
where $m$ is given by Eq.(2.9).

\vskip1pc
\noindent{ 2.2 The Neighborhood of Singularities}
\vskip1pc

In paper I, as a lemma to the positive quasi-local mass theorem,
we proved that when the weak energy condition $\rho\ge 0$
holds, $R^{'2}\le 1$ everywhere in any
regular geometry. Thus if $R^{\prime2}>1$ anywhere the geometry must be
singular. Let us suppose that $R^{\prime2}>1$ at some point.
Then,  when $K_{ab}$ satisfies Eq.(2.3) and
$\alpha> 0.5$, Eq.(2.1b) implies that $R^{\prime\prime} <0$,
so that $R^\prime$ is decreasing there. This can only occur
by $R^\prime$ falling through $R^\prime=-1$. Once $R^\prime$
falls below this value it will continue decreasing
monotonically thereafter. The surface $R^\prime=-1$ in the
configuration space therefore acts as a oneway membrane.
Suppose that the circumferential radius is $R_0$
when  $R^\prime=-1$. We know now that the solution must
crash, {\it i.e.} $R\to 0$ in a finite proper distance
which is less than or equal to $R_0$ from that point.
In fact, this is the  only way the spatial geometry can become singular.

How do we know that we have covered all
possible singularities? We argue that the converse of the lemma
holds. In general, $-1< R^\prime \le 1$ if and only if the geometry
is non-singular. As we will see, a singularity with
$R^\prime=-1$ is a result of a very special fine-tuning
of the matter distribution.

In the neighborhood of the point $\ell=\ell_S$ at which
$R=0$, Eq.(2.4) implies that

$$K_R\sim {C_\alpha(\ell_S)\over R^{1+\alpha}}\,,\eqno(2.10)$$
where

$$C_\alpha(\ell)= 4\pi\int_0^{\ell} d\ell JR^{1+\alpha}\eqno(2.11)$$
is finite. $K_R$ will therefore
be singular (for any physically acceptable value of $\alpha$)
if the geometry pinches off unless the current is tuned such that

$$C_\alpha(\ell_S) =0\,.\eqno(2.12)$$
Now, if $K_R\ne0$, and $\alpha>0.5$, the most singular
term in Eq.(2.8$'$) is the quadratic in $K_R$. This implies that

$$R^{\prime2}\sim R^2 K_R^2\eqno(2.13)$$
in the neighborhood of $R=0$, or
$R^{\prime2} \sim C_\alpha^2/  R^{2\alpha}$.
Generically, therefore, $R^{'2}$ diverges.
The solution is

$$R\sim
\left({ C_\alpha\over\alpha+1}\right)^{1\over\alpha+1}
(\ell_S-\ell)^{1\over\alpha+1}
\,.\eqno(2.14)$$
If $\alpha>0.5$, such spatial singularities are more
severe than the strong singularities discussed in paper II
which are consistent with
the Hamiltonian constraint at a moment of time symmetry.
We will refer to the generic kind of singularity
driven by extrinsic curvature as a strong $J$-type singularity.
As $\alpha$ increases, the power law determining the strength of
the singularity increases.
Note that the limit $\alpha\to\infty$ (the polar gauge discussed
in I) is extremely singular. This is, however, a gauge artifact reflecting
how poor the polar gauge really is.

Unlike the strong singularities occurring in MSCs, at which the scalar
curvature ${\cal R}$ remained finite, ${\cal R}$ will generally
blow up (as $ K_R^2 \sim 1/(\ell_S-\ell)^2$). On dimensional grounds, we expect
all curvature scalars to blow up as $1/(\ell_S-\ell)^2$ as we
approach a singularity unless there is some constraint
obstructing them from doing so.

It is important to confirm that $m$
remains suitably bounded as we approach a strong singularity.
We do this by demonstrating that the volume integral (2.9) is always finite.
We note that for suitably bounded $\rho$ and $J$,

$$
(\rho R^2 R^\prime, J R^3 K_R)\sim
(\rho,J) (\ell_S-\ell)^{-\left({\alpha-2\over \alpha+1}\right)}
\,.\eqno(2.15)$$
If $\alpha\le 2$, the integrand itself remains finite.
In general, the integral will be finite if the exponent of
$(\ell_S-\ell)^{-1}$ is bounded by one. But ${(\alpha-2)/(\alpha+1}) < 1$
for all finite values of $\alpha$ thus guaranteeing that the integrals over
$R^2 \rho R^\prime$ and $R^3 J K_R$ converge.

It is clear that $m(\ell_S)$ is always finite. Its
sign, however, will depend on the details of the current flow.
This is obvious from the definition Eq.(2.8).
Even if $R^{\prime2}>1$, a sufficiently large value of $K_R$
can render $m$ positive. In particular, unlike the value of $m$ assumed
at strong $\rho$-singularities of MSCs which is always negative, the sign
can assume either value. Indeed $m$ need never even be
negative in a singular geometry. Though $R^\prime$ decreases
monotonically, $m$ nonetheless remains positive. There is no conflict
with the positive QLM theorem. In our examination of MSCs in
paper II, we found that $m$ is positive everywhere except at the origin
or in a neighborhood of it if and only if
the geometry is non-singular. This is a consequence of the
coincidence of the converse of the bounded $R^{\prime2}$ lemma
and the converse of the positive QLM theorem
when $K_{ab}=0$. In the general case, when $K_{ab}\neq 0$,
no such coincidence occurs.

What are the implications of the integrability condition,
Eq.(2.12)? If Eq.(2.12) is satisfied the strong $J$
singularity is moderated to one which is only strong a la $\rho$.
The behavior in the vicinity of the singularity will then be
determined by the $m/R$ term in Eq.(2.8$'$) even if the system was
originally `driven' towards the singularity by extrinsic
curvature. If, in addition,

$$m(\ell_S)=4\pi\int_0^{\ell_{S}} d\ell \left[
\rho R^2 R^\prime + J R^3 K_R\right] = 0
\,,\eqno(2.16)$$
the singularity will be a weak one with $R^\prime(\ell_S)=-1$.
The corresponding bag of gold will be a regular closed universe.

These integrability conditions depend on $\alpha$. If
a given function $J$ satisfies Eq.(2.12) with one
value of $\alpha$, generally it will not satisfy that condition
with any other value. What is missing is a spacetime diffeomorphism
invariant statement of the integrability.

If $J$ is positive (or negative)
everywhere, $C_\alpha(\ell)$ defined by Eq.(2.11) cannot vanish.
Thus, if matter is collapsing or exploding everywhere,
all singularities must be strong $J$-type singularities.

This contrasts with the obstruction, $\rho^\prime <0$,
discussed in paper II, {\it prohibiting} the formation of any singularity
when $K_{ab}=0$. In general, we note that on performing an integration
by parts on the first term, $m$ can be rewritten

$$m={4\pi\over 3}\rho R^3 + 4\pi \int_0^{\ell} d\ell R^3
\left[J  K_R - \rho^\prime\right]
\,.\eqno(2.17)$$
The first term is manifestly positive. So is the third if
$\rho^\prime\le 0$. If $J$ is positive (negative)
everywhere then so is $m$ in any $\alpha$ - gauge. However,
the third term appearing on the RHS of Eq.(2.8$'$) may still pull the
geometry into a singularity if $J$ is sufficiently large.
The peculiarity of momentarily static configurations with
$\rho^\prime <0$ discussed in paper II can clearly be destabilized by
the motion of matter.

All regular closed cosmologies simultaneously satisfy
two integrability conditions, Eqs.(2.12) and (2.16).
There can be no net flow of material from one pole to the other.
In particular, $J$ must change sign between the poles.
In addition, Eq.(2.17) tells us that

$$ m(\ell_S)=4\pi\int_0^{\ell_{S}} d\ell R^3
\left[J K_R - \rho^\prime \right]=0
\,.\eqno(2.18)$$
In particular, $J K_R - \rho^\prime$ must change sign
between the poles. These conditions will be examined in
the closed cosmological context in a subsequent publication [4].

\vskip1pc
\noindent{ 2.3 No strong $J$ singularities in the Euclidean Theory}
\vskip1pc

The singularity structure we have investigated has
one important consequence for Euclidean general relativity.
If the sign of the quadratic term in $K_R$ appearing in Eq.(2.8) had been
negative, instead of facilitating the occurrance of
singularities it would have presented an obstacle to their occurrance.
Any non-vanishing extrinsic curvature would therefore tend
to stabilize the spatial geometry against singularity formation.
We note that there is
precisely such a sign switch in the Hamiltonian constraint of Euclidean
general relativity. The Bianchi identities there tell us that the solutions
of the constraints represent all possible configurations the system
may assume as it is evolved with respect to Euclidean time.
This suggests that gravitational instantons will tend to
be more regular than their Lorenzian counterparts. In fact, the
most singular Euclidean geometries will occur when the geometry
is  momentarily static. In a tunneling Euclidean
four-geometry, such three-geometries correspond to the initial and final
hypersurfaces of the Lorentzian spacetimes between which it interpolates.
If these hypersurfaces are themselves non-singular, {\it i.e.}
do not involve Planck scale structures, then
Planck Scale physics does not enter the semi-classical description
of tunneling between them. This would appear to validate
the application of the semi-classical approximation.

\vskip2pc
\noindent{\bf 3 SOLUTIONS OF THE CONSTRAINTS AS TRAJECTORIES ON THE
$(\omega_+,\omega_+)$ PLANE}
\vskip1pc
In paper I we found that a very useful representation of the
phase space was provided by the representation of
solutions to the constraints as trajectories on the
optical scalar plane. We have [1,7]

$$\eqalign{
(\omega_+)^\prime= & -8\pi R(\rho-J) -{1\over 4R}
\Big[2\omega^2_+ -4 - 4\omega_+ KR -\omega_+\omega_-\Big]\,,\cr
(\omega_-)^\prime=
&-8\pi R(\rho+J) -{1\over 4R}
\Big[2\omega^2_- -4 + 4\omega_- KR -\omega_+\omega_-\Big]\,.\cr}
\eqno(3.1a,b)$$

\vskip1pc
\noindent { 3.1 Non-Singular Geometries off the support of matter}
\vskip1pc

In this section, we will focus on the
behavior of trajectories outside the support of matter.
We will suppose that the interior solution is regular.
This is a more useful exercise than it might appear at first sight.
This is because, as we have seen, the behavior of trajectories
depends non-locally on the sources, $\rho$ and $J$. In
particular, the appearance of singularities does not
depend sensitively on the values of $\rho$
and $J$ in the immediate vicinity of the singularity. In addition,
as we will see, the behavior of vacuum trajectories upon entering
into a shell of matter is described in a simple way.

Let us first recall briefly the case the momentarily static
solution outside matter. We set $J=0$ everywhere.
We found that

$$R^{\prime2}=1-{2m_\infty\over R} \,.\eqno(3.1)$$
where

$$m_\infty = 4\pi\int_0^{\ell_0} d\ell R^2 \rho R^\prime\,.
\eqno(3.2)$$

If $m_\infty \le 0$ and $R^\prime(\ell_0)\le 0$, the vacuum geometry will
be singular. If $R^\prime(\ell_0) >0$, the vacuum geometry is non-singular
but the positive quasi-local mass theorem tells us that the interior
must harbor a geometrical singularity.

If $m_\infty >0$, there is no way the geometry can be
singular. If $R^\prime(\ell_0)>0$, the solution grows
monotonically. If $R^\prime (\ell_0)\le 0$, $R$ will decrease
until an apparent horizon forms at $R = 2m_{\infty}$ and thereafter increase.

When  $J\ne0$ the solution is considerably less simple.
Let us rewrite Eq.(2.8) and (2.9) as

$$R^{\prime2}=1-V(R)\,,\eqno(3.3)$$
where the potential $V(R)$ is given by

$$V(R)={2m_\infty\over R} - {C^2_\alpha\over R^{2\alpha}}\,,
\eqno(3.4)$$

$$m_\infty =
4\pi \int_0^{\ell_0} d\ell R^2\left[\rho R^\prime
+ J R K_R \right]\,,\eqno(3.5)$$
and $C_\alpha= C_\alpha (\ell_0)$
is given by Eq.(2.11). We suppose that $0.5 <\alpha<\infty$.

If $m_\infty$ is negative and $R^\prime(\ell_{0})\le 0$ then, as before,
the potential is monotonic and unbounded
from below, and the geometry will be singular. It will generally be
a strong $J$-singularity unless $J$ is tuned such that $C_\alpha(\ell_0)=0$.
If $R^\prime(\ell_0) >0$, the vacuum geometry is non-singular
but the positive quasi-local mass theorem tells us that the interior
must be singular.

What is much more interesting is the case $m_\infty>0$. Unlike the case of
a MSC, a positive $m_\infty$ does not guarantee a
non-singular exterior geometry.

If $m_\infty$ is positive, the potential possesses a maximum at the point

$$R_c=\left({\alpha  C_\alpha^2\over m_\infty}\right)^{1\over 2\alpha-1}
\,.\eqno(3.6)$$
The value assumed by the potential at this point is

$$
V(R_c)=(2-\alpha^{-1}){m_\infty\over R_c}\,.\eqno(3.7)$$
There are two possibilities we need to consider:

If $V(R_c)<1$, the geometry
will be singular outside if and only if $R^\prime(\ell_0)<0$.

If $V(R_c)\ge 1$, the nature of the geometry will depend
not on $R^\prime(\ell_0)$ but on the relative values of
$R(\ell_0)$ and $R_c$. If $R(\ell_0)<R_c$
the geometry will always be singular. If $R(\ell_0)>R_c$,
however, there is no way the geometry can be singular.
The qualitative dependence on the sign of $R^\prime(\ell_0)$ is
the same as that for $m_\infty>0$ when $J=0$.

The condition $V(R_c)< 1$ is equivalent to the inequality

$$
m_\infty^{2\alpha} <
(2\alpha-1)\left({\alpha\over 2\alpha -1}\right)^{2\alpha} C_\alpha^2
\,.\eqno(3.8)$$
Typically Eq.(3.8) will hold for a given $m_\infty$
whenever $C_\alpha$ is large which corresponds,
roughly speaking, to a large material current. However,
because $m_\infty$ itself also involves $J$ this criterion is
not very precise. We need to isolate the dependence of $m_\infty$ on $K_R$.
The inequality can then be cast as an inequality between
$R^\prime$ and $RK_R$.
We recall that $C_\alpha = R^{\alpha} RK_R$.
Now Eq.(3.8) can be cast in the form

$$
\left({1\over 2}\right)^{\alpha}
(1-R^{\prime2} + R^2{K_R}^2 )^{\alpha} <
(2\alpha-1)^{1/2}\left({\alpha\over 2\alpha -1}\right)^{\alpha} | RK_R |
\,.\eqno(3.9)$$
We can also represent the inequality, $R < R_c$,
as the exterior of the ellipse

$$
R^{\prime2} + (2\alpha -1) R^2 K_R^2\ =1
\eqno(3.10)$$
on the $(R^\prime, RK_R)$ plane.
The beauty about Eqs.(3.9) and (3.10) is that when they are
cast in terms of the optical scalar variables, $\omega_+$ and
$\omega_-$ they are independent of $R$.

Let us examine Eq.(3.9) in greater detail. We first cast
it in the form

$$
f( R K_R) \le  R^{\prime2} \,,
\eqno(3.11)$$
where we define

$$
f(x) = x^2 -
2 (2\alpha-1)^{1/2\alpha}\left({\alpha\over 2\alpha -1}\right)
|x|^{1/\alpha} +1
\,.\eqno(3.12)$$
We note that $f(x)$ is positive everywhere. In particular, $f(0)= 1$ and

$$
f(\pm 1/\sqrt{2\alpha-1})=0=f^\prime (\pm 1/\sqrt{2\alpha-1})
\,.\eqno(3.13)$$
In the neighborhood of the points $x=\pm 1/\sqrt{2\alpha-1}$,

$$
f^{1/2}(x) \sim \sqrt{2\alpha-1\over \alpha} \left|x \mp
1/\sqrt{2\alpha-1}\right|
\,.\eqno(3.14)$$
As $x\to\infty$, $f^{1/2}(x)\sim x$.
The two branches of the function $R^\prime = \pm f(R K_R)^{1/2}$
correspond to the boundary $V(R_c)=1$. They can be
represented on the $(\omega_+,\omega_-)$ plane (see fig.(3.1))
as the union of arc segments $\{C{\cal P}$, ${\cal P}P$, $P{\cal P}'$,
${\cal P}' D\}$ and
$\{C'{\cal P}$, ${\cal P}Q$, $Q{\cal P}'$, ${\cal P}D'\}$, where
the coordinates of the points
${\cal P}$ and ${\cal P}'$, corresponding to the two mimima
of $f$,  are given repectively by
$(-2/\sqrt{2\alpha-1},2\sqrt{2\alpha-1})$ and
$(2/\sqrt{2\alpha-1},-2\sqrt{2\alpha-1})$. Fig.(3.1) corresponds
to $\alpha=2$.

The ellipse defined by Eq.(3.10) is also
represented on fig.(3.1) for $\alpha=2$.
We note that for each $\alpha$, the points
${\cal P}$ and ${\cal P}^\prime$ both lie on this ellipse. The
inequality, $R < R_c$, is represented by the region on the phase plane
outside the ellipse.

What is this figure telling us?
There is a wedgelike region $\Omega_0$, bounded by the arc segments,
$CQ$, $QD$ indicated on
fig.(3.1) which determines the maximum excursion a vacuum trajectory
can make from its point of departure, $P=(2,2)$, and still return home.
This is a disjoint union of two regions,
one in which $V(R_c) > 1$ and $R > R_c$, the other in which
$V(R_c) \le 1$ and $R'> 0$.

Any trajectory which lies outside $\Omega_0$
on exiting the support of matter is necessarily singular.
This region, likewise, decomposed into a disjoint union, one in which
$V(R_c)< 1$ and $R^\prime <0$, the other in which
$V(R_c)\ge 1$ and $R<R_c$.
We note that these considerations did not rely on
any energy condition, dominant, weak or otherwise.

When the DEC holds, we note that $\Omega_0$ reduces to
a proper subset of the domain, $\Omega$,
introduced in paper I to which all non-singular trajectories are confined
For $\alpha=2$, $\Omega$
is given by the square, $|\omega_\pm|\le 2$. The region,
$\Omega-\Omega_0$ is rendered forbidden outside the support of matter.
In particular, we note that the barriers $\omega_\pm=-2$ are
completely out of bounds. There always exists, however, a suitable
$\rho$ and $J$, which can be added within the
region $\Omega$ so as to render the trajectory straying
into this region non-singular. To see this, consider the addition of
a shell with a source four-vector given by Eq.(4.1) at $\ell=\ell_0$.
Both $\omega_+$ and $\omega_-$ will suffer a discontinuity
at the shell. The discontinuity
($\Delta\omega_\pm$) is given by integrating Eqs.(3.1a) or (3.1b)
across the surface:

$$
\Delta\omega_\pm = -8\pi R(\ell_0)(\sigma \mp j)
\,.\eqno(3.15)$$
By a suitable choice of $\sigma$ and $j$ it is always possible to
raise or reduce one or the other of $\omega_\pm$
while leaving the other unchanged. In particular, as the arrow
on the point ${\cal Q}$ indicates the value of
$\omega_-$ can be reduced in such
a way that the trajectory is delivered back to safety albeit by
flirting dangerously close to the singular point $Q$.

What is the physical significance of the points
${\cal P}$ and ${\cal P}^\prime$? These are both fixed points
of $\omega_+$, $\omega_-$ and $R$ outside matter:

$$
\omega^\prime_+ =0\quad \omega_-^\prime =0\quad
R^\prime =0\,.\eqno(3.16)$$
As a result, $R^{\prime\prime}=0$ and all higher
derivatives vanish at these points. If we exit matter at any
one of these two points, the spatial geometry
degenerates into an semi-infinite cylinder,
$S^2\times R_+$ outside. They are clearly unstable
fixed points. Under any small perturbation, the
vacuum trajectories terminating on either of these points
will find themselves either returned to the origin, $P$
or consigned to singular oblivion.

The radius of this cylinder is fixed by the value of the ADM mass.
We note that Eq.(3.8) implies

$$R_0 =\Bigg( 2 - {1\over \alpha}\Bigg) m_\infty\,.\eqno(3.17)$$

We have sketched the exterior behavior explicitly for $\alpha=2$ on
fig.(3.1). How sensitively dependent is this picture on the
gauge parameter, $\alpha$?

If $\alpha$ is reduced below two,
the points ${\cal P}$ and ${\cal P}^\prime$
slide out along the $R^\prime=0$ diagonal in opposite
directions reaching infinity at $\alpha=0.5$ ---
the superspace lightcone value.
If $\alpha>2$, ${\cal P}$ and ${\cal P}^\prime$ converge on the
$R^\prime=0$ diagonal, coinciding asymptotically on the
$K_R=0$ axis as $\alpha\to \infty$, the polar gauge value
(discussed in paper I). We see explicitly how
polar gauge imitates a moment of time symmetry in a
very singular way.

When $\alpha=1$, Eq.(3.9) reads
$ R^{\prime2} > (RK_R \mp 1)^2$,
where the $\mp$ correspond respectively to $|K_R|=\pm K_R$.
These are two cones with apices at the points ${\cal P}$
and ${\cal P}^\prime$ given by $RK_R=\pm 1$, $R^\prime=0$.
$\Omega_0$ is now simply a square, whereas $\Omega$ is some more complicated
figure.

We note, however, that while the partition of the $(\omega_+,\omega_-)$
plane depends qualitatively on $\alpha$,
topologically it is identical
to the partition illustrated in fig.(3.1) for $\alpha=2$.

In the vacuum region the trajectory is determined by two
constants, $m_\infty$ and $C_{\alpha}$, which are related to
the optical scalars via
$$\omega_+\omega_- = 4 - {8m_\infty \over R}\,,\eqno(3.18)$$
and
$$\omega_+ - \omega_- = 4RK_R = {4C_\alpha \over R^\alpha}\,.
\eqno(3.19)$$

There are two possible solutions
$$\eqalign{
\omega_\pm =& {2C_\alpha \over R^\alpha}\big(\pm 1 + \sqrt{1 +
{R^{2\alpha - 1}(R - 2m_\infty) \over C^2_\alpha}}\big)\,,\cr
\omega_+ + \omega_- =&{4C_\alpha \over R^\alpha} \sqrt{1 +
{R^{2\alpha - 1}(R - 2m_\infty) \over C^2_\alpha}}\,.\cr
} \eqno(3.20a)$$
and
$$\eqalign{
\omega_\pm =& {2C_\alpha \over R^\alpha}\big(\pm 1 - \sqrt{1 +
{R^{2\alpha - 1}(R - 2m_\infty) \over C^2_\alpha}}\big)\,,\cr
\omega_+ + \omega_- =&-{4C_\alpha \over R^\alpha} \sqrt{1 +
{R^{2\alpha - 1}(R - 2m_\infty) \over C^2_\alpha}}\,.\cr
} \eqno(3.20b)$$

{}From these equations it is easy to analyse the behaviour of
a trajectory as it approaches a singularity. Let us assume that
$C_\alpha > 0$. Since we approach a singularity we can assume
that $R$ is small and positive and that $4R' = \omega_+ +
\omega_-$ is negative. This means that the trajectory is given
by Eq.(3.21b). Further, since $\alpha > 0.5$ the second term in
the square root is much less than 1 and we can use the Taylor
expansion to get

$$\eqalign{
\omega_+ \sim &
{2m_\infty R^{\alpha - 1} \over C_\alpha}\,,\cr
\omega_- \sim & {-4C_\alpha \over R^\alpha}
\,.\cr}\eqno(3.21)$$

Thus we see again the same structure that we described in
Section 2 whereby the singularities we get when $C_\alpha \ne
0$ are stronger than in the MSC configuration.

\vskip1pc
\noindent{\bf 4 THE SPHERICALLY SYMMETRIC SHELL WHEN $J\ne 0$}
\vskip1pc

Clearly we cannot solve Eq.(2.4) exactly. Furthermore, if $J\ne 0$, even the
uniform current/density model becomes non-trivial. The only model
we will solve exactly is the shell.
The dynamics of moving shells is a subject which has received
extensive study. Our focus of interest will, however, be restricted to
an examination of the constraints and the identification of
constraints on the sources avoiding singularities.
{}From one point of view, we have already essentially
solved the problem in our examination of the exterior solution.
This is because all of the interesting physics occurs in this
exterior region.

In paper II, we examined the corresponding MSC. A very rich configuration
space is revealed when we relax $J=0$. Let

$$\eqalign{\rho=&\sigma \delta (\ell-\ell_0)\cr
J=& j\delta (\ell-\ell_0)\,,\cr}\eqno(4.1)$$
where

$$\sigma=\left(\sigma_0^2 + j^2\right)^{1/2}\,.$$
In this form, $\sigma_0$ is the rest mass of the shell. If
$\sigma_0$ is real we satisfy the DEC.

Inside the shell, the space is flat so that $R=\ell$.
As was the case at a moment of time symmetry the
material energy $M$ coincides with its Newtonian value

$$M=4\pi\sigma \ell_0^2\,,\eqno(4.2)$$
and is unaffected by the motion of the shell.

In any $\alpha$-gauge, the momentum constraint implies

$$K_R=\cases{0& $\ell <\ell_0$\cr
	     4\pi \left({\ell_0\over R}\right)^{1+\alpha} j &
$\ell\ge \ell_0\,.$\cr}\eqno(4.3)$$
In particular,

$$K_R(\ell_{0+})= 4\pi j\eqno(4.4)$$
is independent of $\alpha$.
$K_R$ is finite everywhere so long as $R$ remains bounded from below
outside the shell. We can now integrate Eq.(2.1) across
$\ell=\ell_0$ to determine the discontinuity in $R^\prime$ at the shell

$$\Delta R^\prime = -4\pi\sigma \ell_0\,.\eqno(4.5)$$
This discontinuity is independent of $J$.
Eq.(4.5) implies that on the outer surface of the shell,

$$\eqalign{R^\prime(\ell_{0+}) =&1 - 4\pi \sigma\ell_0\cr
=& 1 - {M\over \ell_0} \,.\cr}\eqno(4.6)$$
If $\sigma_0>0$, $R^\prime(\ell_{0+})$ is bounded above by one.
The ADM mass is now given by

$$
m_\infty=m(\ell_{0+})={\ell_0\over 2}\Big(1 -R^{\prime2}(\ell_{0+})\Big)
+{1\over 2}\ell_0^3 K_R^2(\ell_{0+})\,.\eqno(4.7)$$
We now substitute  Eqs.(4.4) and (4.6) into Eq.(4.7). Exploiting
Eq.(4.2), we can express $m_\infty$ in terms of $M$, $\ell_0$,
$\sigma$ and $j$ as follows,

$$m_\infty=M-{M^2\over 2\ell_0} (1- v^2)\,,\eqno(4.8)$$
where we introduce the notation

$$v={j\over \sigma}\,.$$
We note that

$$1-v^2 = {\sigma_0^2\over \sigma^2_0+j^2}$$
is manifestly positive if the DEC holds. We note that
$\alpha$ does not appear in Eq.(4.8). Eq.(4.8) generalizes
Eq.(7.6) of paper II. The binding energy defined in I,

$$-E_B=M-m={M^2\over 2\ell_0}(1-v^2)\,,\eqno(4.8^\prime)$$
is diminished below its Newtonian value
by the motion of the shell. It is, however, still negative
whenever matter satisfies the DEC,
consistent with our hopes and allaying our fears.
If $\sigma_0=0$ corresponding to a null shell
which saturates the DEC
(moving either inward or outward), $E_B=0$.

The divergence of outward bound future (past) directed
null geodesics at the surface of the shell is given by

$$(\Theta_\pm)_0 = {2\over R} \Big(1- 4\pi\ell_0\sigma
(1\mp v) )\Big)\,.\eqno(4.9)$$
We see that $(\Theta_+)_0 < 0$ and hence that a future
horizon must form at some point outside the shell whenever

$$4\pi\sigma \ell_0 (1-v)\ge 1\,.\eqno(4.10)$$
If $v=1$, corresponding to a null outward moving
shell, no future horizon can form.\footnote *
{This does not, however, mean that the geometry
cannot be singular.}
As one would expect it is easier to form an horizon when $v$ is negative.

If $\Delta R^\prime\le -2$ the geometry will be singular. This reads

$${M\over 2\ell_0} \ge 1\,,$$
which is independent of $v$.

Let us interpret the exterior solution we examined in sect.3
in terms of the parameters of the shell model.
We note that $m_\infty$ is given by Eq.(4.8) and

$$\eqalign{C_\alpha=&4\pi \ell_0^{1+\alpha} j\cr
	    =&\ell_0^{\alpha-1} M v\,.\cr}\eqno(4.11)$$
If $m_\infty$ is negative, the positive mass theorem tells us that
the geometry must be singular. We note that $m_\infty \le 0$
in Eq.(4.8) implies

$${M\over 2\ell_0} \ge {1\over 1-v^2}\,.\eqno(4.12)$$
The potential $V(R)$ is monotonic and unbounded
from below. We note that Eq.(4.12) implies that
$R^\prime(\ell_{0+}) < -1$ (see Eq.(4.6)).\footnote {$\dagger$}
{It will be a strong $J$-singularity
unless $v=0$ which is the only way that $C$ can vanish.
The simplest model with a non-trivially vanishing $C$
consists of two shells moving in opposite directions. }

If $m_\infty$ is positive, $V(R)$ possesses a maximum at the point
$R_c$ given by Eq.(3.6). There are two factors
which determine the nature of the exterior
geometry. In sect.3 we saw that the condition $V(R_c)<1$ can
be cast in the form (3.8). Substituting Eq.(4.8) for $m_\infty$
and (4.11) for $C_\alpha$, we get

$$\left(1- {y\over 2}(1-v^2)\right)^{2\alpha} \le
(2\alpha-1) \left({\alpha\over 2\alpha-1}\right)^{2\alpha}
v^2 y^{2(1-\alpha)}\,,\eqno(4.13)$$
where $y:= M/\ell_0$. This relationship clearly depends
on the choice of $\alpha$. When $\alpha=1$, the condition
$V(R_c)\le 1$ is simple:

$${M\over 2\ell_0} > {1\over 1+|v|}\,.$$
If, in addition, $R^\prime(\ell_{0+}) <0$ the geometry will be singular.
Eq.(4.6) then implies

$${M\over \ell_0} \ge 1\,.\eqno(4.14)$$
When the DEC is satisfied all such geometries are singular.

We also note that $\ell_0< R_c$ is equivalent to

$$ {M\over 2\ell_0}> {1\over 1+(2\alpha-1)v^2}\,.$$
When $\alpha=1$,

$$ {M\over 2\ell_0}> {1\over 1+v^2}\,.$$
But, when the DEC holds,

$${1\over 1+v^2}\ge {1\over 1+|v|}\,.$$
There are therefore no geometries which simulataneously satisfy
$V(R_c)\ge 1$ and $\ell_0 <R_c$ when the DEC is satisfied.\footnote *
{Equality $\ell_0= R_c$ gives $R^\prime=0$ outside so that
the exterior of the shell is a cylinder of radius $\ell_0$.
This is singular.} Thus all geometries with $V(R_c)\ge 1$ are non-singular
when the DEC holds. We illustrate the situation in fig.(4.1).
We are now in a position to conclude that

$$M \le  {2\ell_0\over 1+|v|}\,.\eqno(4.15)$$
in any non-singular shell geometry.

Later, we will discover that the (weaker) bound,

$$M\le 2\ell_0\,,\eqno(4.16)$$
holds in a non-singular geometry regardless of $J$ and as a result
also of the DEC. We recall that the shell saturated
this condition when $v=0$ (the MSC result).

It is useful to recall the moment of time symmetry
analysis. It was conjectured by ADM and subsequently
proven by BM\'OM that [8] (also see ref.[2])

$$M-m\ge {M^2/ 2\ell_0}\,.\eqno(4.17)$$
The conjecture was motivated by the fact that in Newtonian gravity, the
configuration that minimized the binding energy for a given total
$M$ is the shell. If (4.17) holds, then Eq.(4.16) follows
by the positivity of the quasilocal mass in any non-singular geometry.
It is important that the bound (4.16) is saturated by a shell.

When $v\ne 0$, Eq.(4.15) is stronger than (4.16).
When $J\ne 0$, we would expect the analogue of Eq.(4.17)
to imply a bound on $M$ at least as good as (4.16).

What is this analogue? If we were to take
Eq.(4.8$^\prime$) at face value, we would conjecture that

$$M-m\ge {M^2 - P^2 \over 2\ell_0}\,,\eqno(4.18)$$
where $P = 4 \pi \int jR^2 dl = 4\pi jl_0^2$.  The positivity of
$m$, however, would now imply that

$$M(1 - P^2/M^2) \le 2\ell_0\,,\eqno(4.19)$$
which is considerably weaker than Eq.(4.16).
The problem is that in the shell, when $v$ is large
the only solutions with $m$ simultaneously small
and positive are all singular (see fig.(4.1).
To be more precise, if we substitute the inequality
Eq.(4.15) into Eq.(4.8$^\prime$) we get

$$m\ge M|v|\,.$$
Hence $m$ is bounded from below if $v$ is non-vanishing.

This suggests that we can do better than (4.16). In fact,
Eq.(4.15) suggests that

$$M +|P|  < 2\ell\,.$$

\vfill\eject

\centerline{\bf THE GLOBAL CHARACTERIZATION }
\centerline{\bf OF TRAPPED SURFACES AND SINGULARITIES}

At a moment of time symmetry, there is a remarkable similarity
between the signal for the presence of an apparent horizon,
$R^\prime=0$  and that for the presence of a singularity,
$R=0$. In paper II, this meant that the techniques which were good for
analyzing apparent horizons were almost always also good
for singularities, and the effort required almost identical.
In general, however, the signal for an apparent horizon
will involve the extrinsic curvature of the spatial hypersurface
and we need to distinguish between future and past horizons.
It is this feature which complicates the analysis of apparent horizons.
It is remarkable that the non-triviality of the momentum constraint
and its coupling to the Hamiltonian constraint does not present
a serious obstacle.

\vskip1pc
\noindent{\bf 5 SUFFICIENCY}
\vskip1pc

In paper II, we demonstrated that sufficiency conditions for
the presence of trapped surfaces and singularities at a moment of
time symmetry could be cast in terms of inequalities of the form,
if $M > $ some constant times $\ell_0$, the geometry must contain a
trapped surface for one constant and a singularity for some other constant.
In this section, we generalize the inequalities of this form
to general initial data.

As we will see, the natural generalizations of $M$ which arise  are
the quantities $M\pm P$, where

$$
P:= 4\pi \int _0^\ell d\ell R^2 J
\,,\eqno(5.1)$$
is a measure of the total current. We note that when the DEC is
satisfied, then

$$
M\pm P = 4\pi \int _0^\ell d\ell R^2 (\rho\pm J)
\eqno(5.2)$$
is positive. A few years ago, Bizon, Malec and \'O Murchadha demonstrated
that if

$$
M- P >{7\over 6}\ell
\,,\eqno(5.3)$$
on a maximal slice, assuming only that $\rho\ge 0$, the spatial
geometry must contain a future trapped surface [5,6]. This
generalizes  the inequality which is valid at a moment of time
symmetry.  They also showed that the numerical coefficient
appearing on the RHS is sharp. They did this by
constructing a solution with $M-P\ge (7/6 -\epsilon)\ell$ but without
any trapped surface. This solution notably did not satisfy the DEC.
More recently, Malec and \'O Murchadha were able to prove that if the DEC
holds, the improved bound,

$$
M-P > \ell
\,,\eqno(5.4)$$
holds [7]. They did this exploiting in a striking way their
reformulation  of the constraints in terms of the optical scalar
variables. The inequality (5.4) is particularly impressive
because it coincides  with the MSC result when $P=0$.

Unfortunately, neither of the inequalities (5.3) and (5.4)
involves spacetime scalars on the LHS so it is not clear what invariant
significance they possess.
Does a change of foliation change these results? To examine this
question, in this section we will examine the issue within the
framework of the one parameter family $\alpha$ foliations. We will
inequalities similar to (5.3) and (5.4) which are not tied
to maximal slicing but are valid in any valid $\alpha$-gauge.

\vskip1pc
\noindent{ 5.1 Trapped Surfaces: Weak Energy, $\alpha$ - Gauge}
\vskip1pc

Let us first assume that only the weak energy condition is satisfied
but instead of considering only $\alpha=2$ as BM\'OM did in their
derivation of Eq.(5.3), we will
suppose that $0.5 <\alpha < \infty$. When $\alpha=2$,
Eq.(5.3) is satisfied. We will prove that, in general,

$$
M -P \ge f(\alpha)\ell
\,,\eqno(5.5)$$
where

$$
f(\alpha) :=1+{1\over2}{(1-\alpha)^2\over 2\alpha-1}
\,,\eqno(5.6)$$
reproducing the bound Eq.(5.3) when $\alpha=2$. However, the minimum of
$f(\alpha)$ is assumed when $\alpha=1$ where we reproduce Eq.(5.4).
Curiously, the gauge providing the best bound when we do not assume
dominant energy is not maximal slicing. The likely reason for this
is that in this gauge, $K_R= P/R^2$.

The original proof by BM\'OM exploited conformal coordinates. Our approach
eschews tying ourselves to any particular spatial coordinate. Not only is
the  end result independent of the spatial coordinate, it is clear that
the  coordinate invariant approach is not only more transparent
but also more efficient.

When Eq.(2.3) holds we can rewrite Eq.(2.1a) in the form

$$
4\pi  \rho R^2  + \partial_\ell\big(RR^\prime\big)=
{1\over 2}\Big(1+(R^\prime)^2\Big)+{1-2\alpha \over 2}R^2 K_R^2
\,.\eqno(5.7)$$
We integrate from $\ell=0$ up to the surface value $\ell$:

$$
M +RR^\prime=  \Gamma +
{1-2\alpha \over 2} \int_0^{\ell} d\ell R^2 K_R^2
\,,\eqno(5.8)$$
where $\Gamma$ is defined (as in paper II) by

$$
\Gamma:={1\over 2}\int_0^{\ell} d\ell
\Big[1+(R^\prime)^2\Big]
\,.\eqno(5.9)$$
We now eliminate $R^\prime$ in the surface term
in favor of the optical scalar $\omega_+$
and $K_R$ using the defining relation (2.6a).
The vanishing of $\omega_+$ signals that the geometry
possesses an apparent horizon. Let us assume that the surface is not
trapped so that $\omega_+>0$.

To eliminate the $K_R$ dependence on the boundary which comes along
with the replacement of $R^\prime$ by $\omega_+$ in Eq.(2.6a),
we note that we can integrate the momentum constraint,
Eq.(I.3.2) to obtain\footnote *
{The privileged role of the gauge with $\alpha=1$ is evident.}

$$
R^2  K_R=  P
+(1-\alpha)\int_0^\ell d\ell R R^\prime  K_R
\,.\eqno(5.10)$$
Substituting Eq.(2.6a) and (5.10) into (5.8) we now obtain

$$
M- P +2\omega_+ =
\Gamma +{1\over 2}(1-2\alpha)\int _0^\ell d\ell R^2 K_R^2
-(1-\alpha)\int _0^\ell d\ell R R^\prime  K_R
\,.\eqno(5.11)$$
When $\alpha>0.5$, the second term on the RHS is manifestly negative.
As such we could discard it to cast (5.11) as an
inequality. However, it is clear that we can do better
by first completing the square in the sum of the second and third terms
before discarding:

$$
\eqalign{{1\over 2}(1-2\alpha)&\int _0^\ell d\ell R^2 K_R^2
 -(1-\alpha)\int _0^\ell d\ell R R^\prime  K_R\cr
&= {1\over 2}(1-2\alpha)\int _0^\ell d\ell \Big(R K_R
+{1-\alpha\over 1-2\alpha} R^\prime \Big)^2
+{1\over2}{(1-\alpha)^2\over 2\alpha-1}
\int_0^{\ell}d\ell R^{\prime2}\cr
&\le  {1\over2}{(1-\alpha)^2\over 2\alpha-1}
\ell \,.\cr}
\eqno(5.12)$$
On the last line, we have used the fact that when $\rho$ is
positive and $\alpha\ge 0.5$, $R^{\prime 2}\le 1$. In addition, under these
conditions, we obtain the same upper bound on $\Gamma$,

$$
\Gamma \le \ell
\,.\eqno(5.13)$$
as we obtained at a MSC. We conclude that if the spherical surface
is not trapped, then

$$
M -P < \left[1+{1\over2}{(1-\alpha)^2\over 2\alpha-1}\right]\ell
\,.\eqno(5.14)$$
Thus if the surface is trapped or the interior contains a trapped
surface Eq.(5.5) holds.

Let us now examine some extreme cases:

We note that $f$ diverges as we
approach the minisuperspace lightcone, $\alpha\to 0.5$ and $\alpha\to \infty$.
While it is tempting to interpret this as a signal of the breakdown of the
gauge on the lightcone, it is also clear that the discarded negative term
blows up at these two values.

Let us consider the two extreme distributions
saturating the DEC everywhere, $P=\pm M$ which are respectively
the cases of a radially outward and a radially inward
moving null fluid. In the former case, Eq.(5.14) becomes a vacuous statement
--- even though we do expect it to be more difficult (if not impossible) to
form an apparent horizon. In the later case, we have that if
$2M \ge f(\alpha)\ell$,
the geometry will possess a trapped surface. It is twice as easy to form
an apparent horizon with an inflowing null fluid as it is with a
stationary fluid.

It is also possible to tighten the sufficiency condition in the same way we
did for MSCs when $\rho^\prime\le 0$ if, in addition,
$J$ has a fixed sign.

We note that in same way we did when $K_{ab}=0$,
when $\rho^\prime\le 0$ we can replace Eq.(5.13) by

$$
\Gamma \le \ell- {M\over 3} -
{8\pi \over R}\int_0^\ell d\ell R^3 J K_R-
\int_0^\ell d\ell R^2 K_R^2
\,.\eqno(5.15)$$
If $J$ is positive (negative) everywhere, the inequality
still holds when the third term on the RHS is dropped.
The (negative) last term on the RHS can now be added to the
(negative) term of the same form in Eq.(5.11)
before the completion of the square. We get

$$
{4M\over 3}- P
< \Big[ 1+ {(1-\alpha)^2\over 2(2\alpha+1)}\Big]\ell
\,.\eqno(5.16)$$
As before, this is minimized when $\alpha=1$ and when $P=0$
again reproduces the result at a moment of time symmetry. We note
that both the LHS and the RHS have been improved. When $\alpha=0.5$,
unlike Eq.(5.14) the RHS of Eq.(5.16) does not diverge.
{}From one point of view, Eq.(5.18) is not
very satisfactory --- we have broken the symmetry between
$J$ and $\rho$. However, on the other hand this asymmetry permits us to
write down a non-vacuous sufficiency condition when the
the spatially averaged DEC is saturated with  $P=M$. Whereas
Eq.(5.14) is vacuous under these conditions, Eq.(5.18) 	provides
the non-trivial statement: suppose $\rho^\prime\le 0$ and the motion of
matter is outward and null, then if

$$
M > 3  \Big[ 1+ {(1-\alpha)^2\over 2(2\alpha+1)}\Big]\ell
\,,\eqno(5.17)$$
the spatial geometry will possess an apparent horizon. When $\alpha=1$,
this value of $M$ is three times larger than that of a
corresponding stationary distribution of matter.

\vskip1pc
\noindent{ 5.2 Trapped Surfaces: Dominant Energy, $\alpha$ - Gauge}
\vskip1pc

When the DEC holds, our experience suggests that the appropriate
variables are the  optical scalars. The optical scalar which
marks the presence of a future trapped surface is $\omega_+$.
Remarkably, only the constraint (3.1a) determining the spatial
derivative of $\omega_+$ will play a role in the determination of the
inequality. Let us first recast Eq.(3.1a) as an equation for
the spatial derivative of $R\omega_+$:

$$
(R\omega_+)^\prime = -8\pi R^2 (\rho- J) +
{1\over 4}\left[2\omega_+\omega_- +4 + 4 RK \omega_+ -
\omega^2_+\right]
\,.\eqno(5.18)$$
This equation can be integrated up to give

$$
R\omega_+ = - 2(M- P) + 2 \Gamma_+
\,,\eqno(5.19)$$
where

$$
\Gamma_+ :=
{1\over 8}\int_0^\ell d\ell \left[2\omega_+\omega_- +4 + 4 RK \omega_+ -
\omega^2_+\right]
\eqno(5.20)$$
are the natural optical scalar generalizations of $\Gamma$.
In particular, when $K_{ab}=0$, $\Gamma_+ =\Gamma$.
In general, in the gauge (2.3),

$$
RK =(2-\alpha) R K_R =(2-\alpha){1\over 4} (\omega_+-\omega_-)
\,,\eqno(5.21)$$
so that

$$
\Gamma_+ =
{1\over 8}\int_0^\ell d\ell \left[4+
\alpha\omega_+\omega_- + (1-\alpha) \omega_+^2 \right]
\,.\eqno(5.22)$$
Let us examine two special cases. The case examined by M\'OM
was $\alpha=2$. Now

$$
\Gamma_+ =
{1\over 8}\int_0^\ell d\ell \left[4+
2\omega_+\omega_-  - \omega_+^2\right]
\,.\eqno(5.23)$$
We note that $0\le (\omega_+-\omega_-)^2$ implies

$$
2\omega_+\omega_-  - \omega_+^2 \le \omega_-^2
\,.\eqno(5.24)$$
We now exploit the inequality (2.19) to obtain
$\Gamma_+ \le \ell$.
If the surface is not future trapped then $\omega_+ >0$ in Eq.(5.19)
reproducing Eq.(5.4).
If $\alpha=1$, then\footnote * {Note that in this gauge, $\Gamma_+=\Gamma_-$.}

$$
\Gamma_+ =
{1\over 8}\int_0^\ell d\ell \left[4+
\omega_+\omega_- \right] \le \ell
\,,\eqno(5.25)$$
using Eq.(2.22), which is the same as the $\alpha=2$ value.
The $\alpha=1$ bound does not improve even though the energy
condition is more stringent. This suggests that this bound is sharp.

What if $\alpha$ is not one of these two special values? In general, the
bound on both $\omega_+$ and $\omega_-$ depends on $K$. This makes it
less obvious how to bound $\Gamma_+$ for any $\alpha$ other than these
two values. What we can do is bootstrap on Eq.(5.21) to turn
this into a bound which is independent of $K$. Let
$\kappa_R= {\rm Max} |RK_R|$. We note that Eq.(5.21) implies

$$\kappa_R \le {1\over 4}( {\rm Max}\,|\omega_+|
+ {\rm Max}\,|\omega_-|)\,.\eqno(5.26)$$
We know from Eqs.(6.4a \& b) in paper I that
${\rm Max}\,|\omega_+|)={\rm Max}\,|\omega_-|)\le
\kappa + \sqrt{|\kappa|^2 +4}$.
Hence
$$
\eqalign{&\le {1\over 2}(\kappa + (|\kappa|^2 +4)^{1/2})\cr
&={1\over 2} \Big(|2-\alpha|\kappa_R +
(|2-\alpha|^2 \kappa_R^2 + 4)^{1/2}\Big)\,.\cr}
\eqno(5.26')$$
It is straightforward to invert Eq.(5.26$'$) to obtain the bound

$$
\kappa_R\le {1\over \sqrt{1 -|2-\alpha|}}
\,.\eqno(5.27)$$
This bound unfortunately is valid only for $1 <\alpha < 3$.
Eq.(5.21), in turn, implies the bounds on $\omega_+$ and $\omega_-$,

$$
\omega_\pm^2 \le  {4\over 1 -|2-\alpha|}
\,.\eqno(5.28)$$
We note that when $\alpha=2$ we reproduce the bounds, Eq.(2.19).
What is remarkable is that these bounds are independent of $J$.
Note that our knowledge of the bound on $R^{\prime2}$ does
not help (nor should it be expected to help) to improve these inequalities.

We now return to Eq.(5.22). We introduce a parameter $b$, and we rewrite

$$
\Gamma_+ =
{1\over 8}\int_0^\ell d\ell \Big[4+ \Big(\alpha +
(1-\alpha)b\Big)\omega_+\omega_-
+ (1-\alpha) (\omega_+^2 - b \omega_+\omega_-)\Big]
\,.\eqno(5.29)$$
We now complete the square on the last term and use an obvious
modification of Eq.(5.24) to obtain the inequality

$$
\Gamma_+ \le
{1\over 8}\int_0^\ell d\ell \left[4+ \big(\alpha +
(1-\alpha)b\big)\omega_+\omega_-
+ (\alpha-1) {b^2\over 4}\omega_-^2\right]
\,.\eqno(5.30)$$

We can now exploit Eqs.(2.22) and (5.28) to obtain,

$$
\Gamma_+
\le \cases{ {1\over 2}\left[1 + \alpha + (1-\alpha)b + {b^2 \over 4}
\right]\ell\,, & $1\le \alpha \le 2$\cr
{1\over 2}\left[ 1 + \alpha + (1-\alpha)b + {\alpha -1\over 3-\alpha}
{b^2 \over 4}\right]\ell\,, & $2\le \alpha \le 3\,.$\cr}
\eqno(5.31)$$
The idea is to find the $b$ which minimizes the RHS.
For fixed $\alpha$, the RHS is minimized when

$$
b=\cases{2(\alpha-1)&  $1\le \alpha\le 2$\cr
         2(3-\alpha)&  $2\le \alpha\le 3\,.$\cr}
                \eqno(5.32)$$
The corresponding values of $\Gamma_+$ are

$$
\Gamma_+
\le \cases{ {1\over 2}\alpha(3-\alpha)\ell\,,& $1\le \alpha \le 2$\cr
 {1\over 2}\left[ \alpha^2  -3\alpha + 4\right]\ell\,, &
 $2\le \alpha \le 3\,.$\cr}
\eqno(5.33)$$
These bounds reproduce the optimal values obtained when $\alpha=1$ and
$\alpha=2$. In the neighborhood of $\alpha=2$, this bound is an
improvement over the bound (5.5) which does not assume the DEC.
Even though the bound (5.27) diverges both as $\alpha\to 1_+$ and as
$\alpha\to 3_-$, the bounds (5.14) at these limit points nonetheless are
finite, and in the former case as we have just seen even coincides
with its optimal value there. It is not clear how to  extend this
technique outside the range $\alpha=2\pm 1$.

\vskip1pc
\noindent{ 5.3 Singularities }
\vskip1pc

It is not obvious how to import the DEC into
the statement of a sufficiency condition for singularities. What we
have is an obvious generalization of the moment of time symmetry result:
We recall that, in general,

$$
M +RR^\prime=  \Gamma +
{1-2\alpha \over 2} \int_0^{\ell} d\ell R^2 K_R^2
\,.\eqno(5.34)$$
We proceed exactly as for a moment of time symmetry.
$\Gamma$ is always bounded by one whenever $\rho$ is positive.
Furthermore, $R^\prime\le 1$ so that $R(\ell)\le \ell$ everywhere
on a non-singular geometry and $R^\prime \ge -1$. The  surface term
is therefore bounded from below by $-\ell$.
Finally, the second term is negative whenever $\alpha\ge 0.5$ ---
The $K_{ab}$ dependence is trivially handled. Thus  we get

$$
M \le 2\ell \,,\eqno(5.35)$$
independent of the value of $\alpha\ge 0.5$ which is
exactly the result at a moment of time symmetry.

As at a moment of time symmetry, if we place constraints on the sources
it is possible to tighten the inequality. We note that when $\rho^\prime\le0$
and $J$ is positive (or negative) everywhere, then Eq.(5.15) can
be truncated even more brutally, $\Gamma \le \ell- M/3$.
We get

$$
M \le {3\over 2}\ell
\,.\eqno(5.36)$$
Unlike the moment of time symmetry discussion we cannot claim that
this represents a universal bound when $\rho^\prime <0$
and $J$ is positive (negative). The reason is that the geometry can still
turn singular if $J$ is large enough.

We note that there is no obvious way of introducing $P$ into
either Eq.(5.35) or Eq.(5.36). The singularity
condition is not symmetrical in $M$ and $P$.

\vskip1pc
\noindent{\bf 6 NECESSITY}
\vskip1pc

We noted in paper II, in our examination of a moment of time symmetry,
that the necessary conditions we were able to
formulate with respect to $M$ and $\ell$ were extremely weak.
If $J\ne 0$ even these conditions appear to be
beyond our reach. What one
can do is provide generalizations of the necessary conditions
which where formulated with respect to the variables,
$\rho_{\rm Max}$ and $\ell_0$.
These inequalities assumed the form

$$\rho_{\rm Max}\ell^2 < {\rm constant}\,.\eqno(6.1)$$
Typically, we would expect $|J_{\rm Max}|$ and $\alpha$ to enter into
this description. We would expect that by appealing to the DEC
the inequalities should simplify.
Crucial to the derivation of Eq.(6.1) are two
simple Sobolev inequalities of the form
$$
S\int_0^{\ell_1} d\ell \, R^2\le
\int_0^{\ell_1} d\ell \, R^{\prime2}
\,,\eqno(6.2)$$
where $S$ depends on the boundary conditions satisfied by $R$.
In general $R(0)=0$. At the first trapped surface, $R'(\ell_1)=0$
and $S= \pi^2/4\ell_1^2$. At a singularity, $R(\ell_1)=0$ and
$S=\pi^2/\ell_1^2$. At a singularity, we found that
$R$ tends to zero like $R\sim (\ell-\ell_1)^{2/3}$ so that
$R'$ diverges like $(\ell-\ell_1)^{-1/3}$.  Even though
$R'$ diverges so that the integrand on the RHS of Eq.(5.9)
diverges, the integral itself remains finite.
When $J\ne 0$, however, $R$ diverges more strongly,
$R\sim (\ell-\ell_1)^{1/1+\alpha}$ (see Eq.(2.14)) so that
$R'\sim (\ell-\ell_1)^{-\alpha/1 +\alpha}$.
Thus the integral on the RHS of Eq.(5.9) will only exist if
$\alpha<1$ --- outside the range found to provide the best sufficiency
results in Sect.5.1. Thus, whereas we found that we
could optimize the inequalities of necessity at a moment of
time symmetry by weighting $R^{\prime2}$
by an appropriate power of $R$, a non-trivial weighting will be
essential when $J\ne 0$ at least in the case of singularities.

\vskip1pc
\noindent{ A Bound on $K_R$ }
\vskip1pc

To form a necessary condition for singularities
it is important to possess some control over $K_R$
in a manner which does not require the geometry to be regular.
In particular, we cannot exploit Eq.(5.27) which is only true in
regular initial data.  It is, however, simple to obtain
a bound on $K_R$ by $|J_{\rm Max}|$ without making any assumptions
about the regularity of the geometry. We have
that

$$
|K_R|\le {4\pi\over R^{1+\alpha}}\int_0^{\ell_1} d\ell R^{1+\alpha} |J|
\,.\eqno(6.3)$$
In general,

$$
|K_R|\le {4\pi |J_{\rm Max}|\over R^{1+\alpha}}\int_0^{\ell_0}
d\ell R^{1+\alpha}
\,.\eqno(6.4)$$
This is the result we will exploit below.
There are some interesting related inequalities. Suppose that
the DEC holds, and $\alpha=1$. We obtain

$$
|K_R| \le {M\over R^2}
\,.\eqno(6.5)$$
This inequality in turn implies that the proper spatial average
of $|K_R|$ is bounded by the product, $\rho_{\rm Max}\ell_1$:

$$
<|K_R|>\,\,\le 4\pi\,\, \rho_{\rm Max}\ell_1
\,,\eqno(6.7)$$
a pretty result, even if we have not found an application for it.
\vskip1pc
\noindent{A Bound on $R'$}
\vskip1pc
We will also require a bound on $R'$ which does not require the geometry
to be globally regular. To obtain this bound, we note that in any
$\alpha$-gauge, the Hamiltonian constraint Eq.(2.1a) reads

$$RR''= {1\over 2} (1-R'^2) +
{1\over 2} R^2 (1-2\alpha) K_R^2 - 4\pi R^2\rho\,.$$
At the origin, we have $R'=1$. At a singularity we have
$R'<0$ whereas at a globally regular solution we have $R'\to 1$
at infinity. If $R'$ has an interior maximum then
$R''$ vanishes there. Hence at that point we have

$$1-R'^2 = R^2 (2\alpha-1) K_R^2 + 8\pi R^2\rho\,.$$
Thus, if we have a standard $\alpha$-slice, {\it i.e.},
$\alpha>0.5$ and if the source satisfies the weak energy condition,
$\rho\ge0$ we must have

$$1-R'^2 \le 0\,,$$
and therefore at the maximum of $R'$ we must have $R'\le 1$. Therefore
this is a global bound independent of whether the slice is regular
or not.

\vskip1pc
\noindent{ 6.1 Singularities}
\vskip1pc

The most naive generalization of Eq.(6.1)
would be an inequality treating $\rho_{\rm Max}$ and
$|J_{\rm Max}|$ symmetrically, of the form: if

$$
(\rho_{\rm Max}+|J_{\rm Max}|)\ell_0^2
< {\rm c}\,,\eqno(6.8)$$
for some constant $c$, the geometry is regular. However, our
experience examining the approach to singularities suggests that this
is too optimistic. The natural inequality we obtain involves not
$J$ but its square, assuming the form: if

$$
(\rho_{\rm Max}\ell_0^2 +{\rm c}_1(|J_{\rm Max}|\ell_0^2)^2)< {\rm c}_2
\,, \eqno(6.9)$$
where $c_1$ and $c_2$ are two constants, the geometry is regular. Even
if matter satisfies the DEC, once we foliate
extrinsically the symmetry is broken. The value of $J$ plays a more
significant role than the value of $\rho$. This is consistent with
our findings in Sect.2 in our examination of the generic behavior
of the metric in the neighborhood of a singularity in an
$\alpha$-foliation of spacetime. The optical scalar variables
suggest that a more judicious gauge involving some mix of intrinsic
and extrinsic variables might restore the symmetry between
$\rho$ and $J$ we have broken with the $\alpha$-parametrized gauges.

We note that Eq.(2.1) implies

$$
{1\over 2}(1+ R^{\prime2}) = (RR')^\prime +
4\pi\rho R^2 + {1\over 2}(2\alpha-1) R^2 K_R^2
\,.\eqno(6.10)$$
The last term is manifestly positive. Suppose that the geometry
is singular at $\ell=\ell_1$. We cannot simply integrate Eq.(6.10) and
discard the boundary term. First of all, $R^{\prime2}$ is not
integrable on the interval $[0,\ell_1]$ and, secondly, the surface
term $RR'$ does not vanish at the singularity unless $\alpha <1$.

What we need to do is multiply Eq.(6.10) by some (positive) power of
$R$ before integration. The relevant power of $R$ will generally depend on
the value of $\alpha$. To restore the divergence appearing in Eq.(6.10)
we need to perform an integration by parts. We now integrate up to $\ell_1$:

$$
{1\over 2}\int_0^{\ell_1}
d\ell R^a(1 + (2a+1) R^{\prime2}) = R^{1+a} R^\prime\Big|_{\ell_1}
+ 4\pi \int_0^{\ell_1} d\ell \rho R^{2+a} + {1\over 2}(2\alpha-1)
\int_0^{\ell_1} R^{2+a} K_R^2
\,.\eqno(6.11)$$
To discard the boundary term, we require $R^{1+a} R'$ to vanish at the
singularity. This implies that

$$a > \alpha-1
\,.\eqno(6.12)$$
This choice of $a$ simultaneously bounds the integral over
$R^a R^{\prime2}$.

We also will need to place a bound on the last term on the RHS of
Eq.(6.11). We exploit Eq.(6.4) to bound $K_R$.
The problem is that this bound
involves the positive power of $R$, $R^{1+\alpha}$ in the denominator
which is difficult to control. We obtain the bound,

$$
\int_0^{\ell_1} d\ell R^{2+a} K_R^2\le (4\pi)^2  J_{\rm Max}^2
\int_0^{\ell_1} d\ell R^{a - 2\alpha}
\left(\int_0^\ell d\ell R^{1+\alpha} \right)^2
\,,\eqno(6.13)$$
on the term quadratic in $K_R$. If the weighting term
is chosen such that

$$
a\ge 2\alpha
\,,\eqno(6.14)$$
the denominator problem is solved. Fortunately, such values are
consistent with Eq.(6.12) for all physically acceptable values of $\alpha$.
The RHS of Eq.(6.11) is clearly simplest when

$$a=2\alpha\,.\eqno(6.14^\prime)$$
This is the value we will henceforth adopt for $a$.
The expression is still not very useful as it stands.
A remarkable fact, however, is that we can bound it
by an integral over $R^{2(1+\alpha)}$. To understand why this is
important, note that the integral over $R^{\prime2}$ appearing on the
LHS of Eq.(6.11) can be cast in the form

$$\int_0^{\ell_1} d\ell\, R^{2\alpha} R^{\prime2} =
{1\over (\alpha +1)^2}\int_0^{\ell_1}d\ell\, (R^{\alpha  +1})^{\prime2}
\,.\eqno(6.15)$$
The Sobolev inequality can be exploited to place a bound on the
integral over the function $R^{2(1+\alpha)}$:

$$
S_0\int_0^{\ell_1}d\ell\, R^{2(\alpha  +1)} \le
{1\over (\alpha +1)^2}\int_0^{\ell_1}d\ell (R^{\alpha  +1})^{\prime2}
\,,\eqno(6.16)$$
where the constant $S_0 =\pi^2 /\ell_1^2$ is the Sobolev constant which is
relevant for functions which vanish at both $\ell=0$ and $\ell=\ell_1$.

We now prove the existence of a bound of the form

$$
\int_0^{\ell_1} d\ell \left(\int_0^\ell d\ell R^{1+\alpha} \right)^2
\le C \int_0^{\ell_1}d\ell\, R^{2(\alpha  +1)}
\,,\eqno(6.17)$$
for some appropriate constant $C$. A crude bound is provided by the
positivity of the covariance for any power $R^n$:(Holder Inequality),

$$
<R^n>^2\quad \le\quad <R^{2n}>\,,\eqno(6.18)$$
which implies

$$
\left(\int_0^\ell d\ell\, R^n\right)^2\le
\ell \int_0^\ell d\ell R^{2n}
\,,\eqno(6.19)$$
so that

$$\int_0^{\ell_1} d\ell \left(\int_0^\ell d\ell R^{1+\alpha} \right)^2
\le
{\ell_1^2\over 2} \int_0^{\ell_1} d\ell \,R^{2(1+\alpha)}
\,.\eqno(6.20)$$
We can, however, do better. Let

$$
G(\ell) :=    \int_0^\ell d\ell R^n
\,.\eqno(6.21)$$
Now $G(0)$ =0 and $G'(\ell_1) =0$, for all $n\ge 0$. We apply the Sobolev
inequality to $G$ with the appropriate constant

$$
\int_0^{\ell_1} d\ell \, G_n(\ell)^2 \le
\left({2\ell_1\over \pi}\right)^2
\int_0^{\ell_1} d\ell \, R^{2n}\,.\eqno(6.22)$$
so that

$$
\int_0^{\ell_1} d\ell \,\left(\int_0^{\ell} d\ell R^{2(1+\alpha)}\right)^2
\le\left({2\ell_1\over \pi}\right)^2
\int_0^{\ell_1} d\ell \, R^{2(1+\alpha)}
\,.\eqno(6.23)$$
This is better by a factor of $\pi^2/8$ than the estimate (6.20).
The end result is the bound

$$
\int_0^{\ell_1} R^{2(1+\alpha)} K_R^2\le
64 J_{\rm Max}^2 \ell_1^2 \int_0^{\ell_1}\, R^{2(1+\alpha)}
\,,\eqno(6.24)$$
on the third term on the RHS of Eq.(6.11).  We can now write

$$
1 \le  2\left[4\pi \rho_{\rm Max} + 32
(2\alpha-1)  J_{\rm Max}^2 \ell_1^2
-\left({\pi\over \ell_1}\right)^2
{1+4\alpha \over 2(1+\alpha)^2 }\right]
\int_0^{\ell_1} d\ell R^{2(1+\alpha)}\Big/
\int_0^{\ell_1} d\ell R^{2\alpha}
\,.\eqno(6.25)$$
In paper II, we proved that the ratio of integrals appearing on the
RHS can be bounded as follows (Eq.(II 6.3.16)) ($a=2\alpha$)

$$
{\int_0^{\ell_1} R^{2 + a} d\ell \over
\int_0^{\ell_1} R^{a}d\ell}\, \le {1 + a \over 3 + a}{\ell_1}^2 \,,
\eqno(6.26)$$
which implies

$$
{1\over 2} {3 + 2\alpha\over 1 + 2\alpha  }
+ {1+4\alpha \over 2(1+\alpha)^2 }\pi^2
\le 4\pi \rho_{\rm Max}\ell_1^2  +
32 (2\alpha-1) J_{\rm Max}^2 \ell_1^4
\,.\eqno(6.27)$$
If $\alpha=1$,

$${5\pi\over 32}\left[1 + {4\over 3 \pi^2}\right]
\le  \rho_{\rm Max}\ell_1^2 +
{8\over \pi} (J_{\rm Max} \ell_1^2)^2
\,.\eqno(6.28)$$

In Eq.(6.27), it does not make much sense to claim
that one value of $\alpha$ provides a better
bound than another value. Not only is the inequality invalidated if the
quadratic in $J_{\rm Max}$ is dropped, $J_{\rm Max}$ plays a more
decisive role than $\rho_{\rm Max}$ in the inequality (6.27),
appearing as it does through its square in contrast to
$\rho$ which appears linearly. The MS inequality does
not generalize in the obvious linear way.

If the DEC, the inequality
simplifies. For $\alpha=1$ we obtain

$$
{5\over 6}+{5\pi^2 \over 8}
\le  32 \left(\rho_{\rm Max} \ell_1^2 +{\pi\over 16}\right)^2
- {\pi^2 \over 8}
\,,\eqno(6.29)$$
or

$$
{1\over 8}\left[\sqrt{{5\over 3} +{3\over 2}\pi^2}- {\pi\over 2}\right]
\le  \rho_{\rm Max} \ell_1^2
\,.\eqno(6.30)$$
The LHS $\sim 5/16$, which is approximately half as good as
the moment of time symmetry result.

\vskip1pc
\noindent{ 6.2 Apparent Horizons}
\vskip1pc

We note that Eq.(2.1) implies

$$
{1\over 2}(1+ R^{\prime2}) = (R\omega_+ - R^2 K_R)^\prime +
4\pi\rho R^2 + {1\over 2}(2\alpha-1) R^2 K_R^2
\,.\eqno(6.31)$$
Again the third term on the RHS is manifestly positive.
We can integrate Eq.(6.31) up to the first future horizon

$$
\Gamma =- R K_R|_{\ell_1}+  4\pi \int_0^{\ell_1} d\ell R^2 \rho +
{1\over 2}(2\alpha-1) \int_0^{\ell_1} d\ell R^2 K_R^2
\,,\eqno(6.32)$$
where $\Gamma$  is given by Eq.(5.9).
We note that Eq.(6.3) places a bound on $K_R$ in the surface term.
Thus

$$
\Gamma \le  4\pi \Big(\rho_{Max} + |J_{Max}|\Big)\int_0^{\ell_1}
d\ell R^2 +{1\over 2}(2\alpha-1) \int_0^{\ell_1} d\ell R^2 K_R^2
\,.\eqno(6.33)$$
A linear term in $J_{\rm Max}$ appears in the apparent
horizon inequality condition which is not present in the
singularity inequality. This is a reflection of the
different boundary conditions enforced there.

We can exploit a Sobolev inequality to place a bound on the integral
over the interval $(0,\ell_1)$ of the quadratic $R^2$ by the same
integral over the quadratic, $R^{\prime2}$:

$$
S\int_0^{\ell_1} d\ell \, R^2\le
\int_0^{\ell_1} d\ell \, R^{\prime2}
\,.\eqno(6.34)$$
The relevant boundary conditions are

$$
R^\prime + R K_R =0
\eqno(6.35)$$
at $\ell=\ell_1$. The inequality is saturated by the trigonometric function,

$$
R(\ell) = \sin (\gamma \ell)
\,,$$
which also determines the optimal value of $S=\gamma^2$. The boundary
condition, (6.35) determines $\gamma$ to be the lowest solution of the
transcendental equation,

$$
\tan \gamma \ell_1 = -{\gamma\over K_R}
\,.\eqno(6.36)$$
We note that

$$
\gamma \le {\pi\over 2\ell_1}
\eqno(6.37)$$
if $K_R$ is negative with $\gamma \to \pi/2\ell_1$ as $K_R\to 0$ which
is the moment of time
symmetry bound and $\gamma \to \pi/\ell_1$ as $K_R\to +\infty$.

When we attempt to bound the right hand side we run into the
same problem we faced before with the last term.
In addition, however, we must contend with the
surface term.

The same weighting we found worked
before works again. To restore the divergence appearing in Eq.(6.31)
we need to perform an integration by parts. We
integrate up to $\ell_1$:

$$\eqalign{
{1\over 2}\int_0^{\ell_1}
d\ell R^{2\alpha}(1 + (4\alpha+1) R^{\prime2}) = &
-R^{2(1+\alpha)} K_R\Big|_{\ell_1}\cr
&+ 4\pi \int_0^{\ell_1} d\ell \rho R^{2(1+\alpha)}
+ {1\over 2}(2\alpha-1) \int_0^{\ell_1} R^{2(1+\alpha)} K_R^2
\,.\cr}\eqno(6.38)$$
We now exploit Eq.(6.3) to bound the $K_R$ and $K_R^2$ term as follows
For the former,

$$
R^{2(1+\alpha)} K_R\Big|_{\ell_1} \le
4\pi R^{1+\alpha} J_{\rm Max} \int_0^{\ell_1} d\ell
R^{1+\alpha} \,.\eqno(6.39)$$
The weighting process has broken the symmetry under interchange
of $\rho$ and $J$ of the linear terms on the RHS of Eq.(6.38).
For the term quadratic in $K_R$, we again have ((6.12) with $a=2\alpha$)

$$
\int_0^{\ell_1} R^4 K_R^2\le (4\pi)^2  J_{\rm Max}^2
\int_0^{\ell_1} \left(\int_0^\ell d\ell R^2 \right)^2
\,.$$
We again require a bound on the last term by an integral over
$R^{2(1+\alpha)}$. This time, however, the Sobolev constant
is that which is
relevant for functions which vanish at $\ell=0$ but satisfy
Eq.(6.35) at $\ell=\ell_1$, {\it i.e.},
$S=\gamma^2$, where $\gamma$ is given by (6.36) and (6.37).

The crude bound we derived before, (6.20), is
expected to work better this time.
As before, however, we can do better. This time we let

$$
H(\ell) :=    \int_0^\ell d\ell R^n \Big/\int_0^{\ell_1} d\ell R^n
\,.\eqno(6.40)$$
Now $H(0)$ =0 and $H(\ell_1) =1$ for all $n$. We apply the Sobolev
inequality to $H$ with the appropriate constant

$$
\int_0^{\ell_1} d\ell \, H(\ell)^2 \le
\left({2\ell_1\over \pi}\right)^2
\int_0^{\ell_1} d\ell \, R^{2n}
\Big/\Big(\int_0^{\ell_1} d\ell R^n\Big)^2 \,.\eqno(6.41)$$
so that Eq.(6.23) holds exactly as before and we
again obtain the bound (6.24) for the integral over $K_R^2$.
We can now write

$$
\eqalign{1 \le 2\left[4\pi \rho_{\rm Max} +
32(2\alpha-1) J_{\rm Max}^2 \ell_1^2
-\gamma^2
{1+4\alpha \over 2(1+\alpha)^2 }\right]
&\int_0^{\ell_1} R^{2(1+\alpha)}\Big/
\int_0^{\ell_1} d\ell R^{2\alpha} \cr
+8\pi J_{\rm Max}&\ell^{1+\alpha}_1
\int_0^{\ell_1} R^{1+\alpha} \Big/
\int_0^{\ell_1} d\ell R^{2\alpha} \,.\cr}
\eqno(6.42)$$
An upper bound on $\gamma$ in Eq.(6.42) is provided by its $K_R\to\infty$
limit, {\it i.e.}, $\pi/\ell_1$ and the lower limit is zero.
We can again exploit (6.26) to bound the ratio of the integrals in the
first term of (6.42). In the second term it is clear that

$$
\ell^{1+\alpha}_1
\int_0^{\ell_1} R^{1+\alpha} \Big/
\int_0^{\ell_1} d\ell R^{2\alpha} $$
is bounded by $\ell_1^2$ if $\alpha \le 1$. Therefore a necessary
condition for the appearance of a trapped surface at some proper radius
$\ell_1$ is that

$$
4\pi \rho_{\rm Max}\ell_1^2 +
4\pi{3+2\alpha\over 1+2\alpha} J_{\rm Max} \ell_1^2
+32 (2\alpha-1) J_{\rm Max}^2 \ell_1^4\ge {1\over 2}
{3+2\alpha\over 1+2\alpha} \,.\eqno(6.43)$$

\vskip2pc
\noindent{\bf CONCLUSIONS}
\vskip1pc

This paper concludes a series of three papers on the
identification of the configuration space in spherically
symmetric general relativity. We have attempted to provide a
coherent synthesis of two very different ways of looking at the
constraints, one in terms of the traditional metric variables,
the other in terms of the optical scalar variables.
Which description is appropriate depends very much on the
details of the problem under consideration.

A very satisfying representation has emerged of regular closed
solutions as closed bounded trajectories on the $(\omega_+,\omega_-)$
plane. In this representation, $R$ plays a
secondary role. We have performed the analysis explicitly in vacuum.
We will show elsewhere that this plane also
provides a very profitable representation of
$\alpha$-slicings of the Schwarzschild spacetime [9].

We have presented a variety of necessary and sufficient conditions for the
presence of apparent horizons and singularities in the initial data.
This paper is necessarily more open-ended than either paper I or paper II.
It is clear that some of the Sobolev inequalities exploited in sect.6 can
be sharpened. Indeed, the professional will consider
our approach to functional analysis extremely
heuristic. As physicists, however,  we are more interested in the
fact that such bounds can be established than in squeezing them
for better constants.

Where does one go from here?
The obvious challenge is to generalize this work
to non-spherically symmetric geometries. One needs to bear in
mind, however, that our ability
to describe the configuration space in considerable detail has
relied on features of the spherically symmetric problem which we know
do not admit generalizations.

There is still, however, much that needs to be done
before we can claim to understand spherical symmetry.

We need first of all to examine the classical
evolution. Write down the Einstein equations with respect to
the optical scalar variables. Can we cast the
theory in Hamiltonian form? If the value of these
variables in the analysis of the constraints is anything to go by,
one has every reason to expect that they will
throw light on the solution of the dynamical Einstein equations,
both analytically and numerically. Indeed Rendall
has recently exploited these variables to extablish
a global existence result [10].

A physically interesting question that is extremely relevant is the
identification of initial data that potentially might develop
apparent horizons. In principle it should
be possible to do this exploiting in addition to the
constraints the dynamical Einstein equations
evaluated on the initial hypersurface. These equations involve the
pressure of matter though some equation of state.
The scenario which is most susceptible
to collapse is pressureless matter. We should be able to
exploit this condition to formulate neceasary conditions along the
lines developed in sect.6. At the other extreme, a stiff equation of state
would inhibit collapse. Thus such a scenario might provide
a sufficient condition. A successful analysis of this
nature has the promise of putting an analytical handle on
the physics hinted at in Choptuik's numerical simulations
of the collapse of a massless scalar field [11].

Finally, the bounds on the optical scalars are certain to
have profound implications for the canonical quantization of this model
for gravity [12]. We hope to examine this problem in a subsequent publication.

\vfill
\eject

\centerline{\bf Figure Captions}
\vskip2pc
\noindent {\bf fig.(3.1)} Non-singular exterior vacuum solutions on the
$(\omega_+,\omega_-)$ plane for $\alpha=2$.
All non-singular exterior trajectories lie within
the `wedge' shaped region, $\Omega_0$, bounded by the
arc segments, $CQ$ and $QD$.

\vskip1pc
\noindent {\bf fig.(4.1)} $M/2\ell_0$ vs. $v$ in the
shell model. All non-singular geometries lie below the curve
$V(R_c)= 1$.

\vfill\eject
\noindent{\bf REFERENCES}
\vskip1pc

\item{1.} J. Guven and N. \'O Murchadha, gr-qc/9411009 (1994)
This will be referred to, henceforth, as paper I. An
extensive list of references is provided here.
\vskip1pc
\item{2.} J. Guven and N. \'O Murchadha, gr-qc/9411010 (1994)
This will be referred to, henceforth, as paper II.
\vskip1pc
\item{3.} For references, see ref.[1].
\vskip1pc
\item{4.} J. Guven and N. \'O Murchadha, unpublished (1995)
\vskip1pc
\item{5.} P. Bizo\'n, E. Malec and N. \'O Murchadha {\it Phys. Rev. Lett.}
{\bf 61}, 1147 (1988)
\vskip1pc
\item{6.} P. Bizo\'n, E. Malec and N. \'O Murchadha {\it Class Quantum Grav}
{\bf 6}, 961 (1989)
\vskip1pc
\item{7.} E. Malec and N. \'O Murchadha
{\it Phys. Rev.} {\bf D49} 6931 (1994)
\vskip1pc
\item{8.} P. Bizo\'n, E. Malec and N. \'O Murchadha {\it Class Quantum Grav}
{\bf 7}, 1953(1990)
\vskip1pc
\item{9.} J. Guven and N \'O Murchadha, unpublished, (1995)
\vskip1pc
\item{10.} A. Rendall, gr-qc/9411011
\vskip1pc
\item{11.} M. Choptuik, {\it Phys. Rev. Lett.} {\bf 70} 9 (1993)
\vskip1pc
\item{12.} The canonical quantization of vacuum spherically
symmetric general relativity with Schwarzschild topology has been
examined by K. Kuchar, {\it Phys. Rev} {\bf D50} 3961 (1994)

\bye